\documentclass{emulateapj}
\usepackage{amsmath}
\usepackage{natbib}
\usepackage{float}
\usepackage{subfigure}
\usepackage{xcolor}
\usepackage{mathrsfs}
\usepackage[dvipdfm]{hyperref}

\def\medd{{\dot M_{Edd}}}

\def\muunit{\rm g~cm^{-1}~{s}^{-1}}
\def\be{\begin{equation}}
\def\ee{\end{equation}}
\catcode`\@=11 
\def\@versim#1#2{\vcenter{\offinterlineskip
        \ialign{$\m@th#1\hfil##\hfil$\crcr#2\crcr\sim\crcr } }}

\shorttitle{\emph{Fermi} bubbles inflated by winds from accretion flow \uppercase\expandafter{\romannumeral2}}
\shortauthors{Mou et al}

\begin{document}
\bibliographystyle{apj}
\title
{The accretion wind model of the \emph{Fermi} bubbles (II): radiation}

\author {Guobin Mou \altaffilmark{1,2}, Feng Yuan\altaffilmark{1}, Zhaoming Gan\altaffilmark{1}, Mouyuan Sun\altaffilmark{3,4}}
 \altaffiltext{1}{Shanghai Astronomical Observatory, Chinese Academy of Sciences, 80Nandan Road, Shanghai, China; fyuan@shao.ac.cn (FY)}
 \altaffiltext{2}{University of Chinese Academy of Sciences, 19A Yuquan Road, 100049, Beijing, China}
 \altaffiltext{3}{Department of Astronomy \& Astrophysics and Institute for Gravitation and the Cosmos, 525 Davey Lab, The Pennsylvania State University, University Park, PA 16802, USA}
 \altaffiltext{4}{Department of Astronomy and Institute of Theoretical Physics and Astrophysics, Xiamen University, Xiamen, Fujian 361005, China}

\begin{abstract}
  In a previous work, we have shown that the formation of the \emph{Fermi} bubbles can be due to the interaction between winds launched from the hot accretion flow in Sgr A* and the interstellar medium (ISM). In that work, we focus only on  the morphology. In this paper we continue our study by calculating the gamma-ray radiation. Some cosmic ray protons (CRp) and electrons must be contained in the winds, which are likely formed by physical processes such as magnetic reconnection. We have performed MHD simulations to study the spatial distribution of CRp, considering the advection and diffusion of CRp in the presence of magnetic field. We find that  a permeated zone is formed just outside of the contact discontinuity between winds and ISM, where the collisions between CRp and thermal nuclei mainly occur. The decay of neutral pions generated in the collisions, combined with the inverse Compton scattering of background soft photons by the secondary leptons generated in the collisions and primary CR electrons can well explain the observed gamma-ray spectral energy distribution. Other features such as the uniform surface brightness along the latitude and the boundary width of the bubbles are also explained. The advantage of this ``accretion wind'' model is that the adopted wind properties come from the detailed small scale MHD numerical simulation of accretion flows and the value of mass accretion rate has independent observational evidences. The success of the  model suggests that we may seriously consider the possibility that cavities and bubbles observed in other contexts such as galaxy clusters may be formed by winds rather than jets.
\end{abstract}


\section{INTRODUCTION}

Two giant gamma-ray bubbles above and below the Galactic plane were discovered by the {\it Fermi Gamma-ray Space Telescope} (\citealt{Su10}; \citealt{Dobler2010}). Subsequently, additional observations have been performed and abundant data has been accumulated (\citealt{Su2012}; \citealt{Hooper2013}; \citealt{YangRZ2014}; \citealt{Ackermann2014}). The main observational results are summarized as follows.

\begin{itemize}
\item The bubbles extend to $\sim 50^{\circ}$ above and below the Galactic plane, and the width is $\sim 40^{\circ}$ in longitude.
\item The surface brightness is roughly uniform, with a significantly enhanced substructure in the south bubble, which is called the ``cocoon''.
\item {The boundary of the bubbles is sharp, with a width of about $3^{\circ}$.}
\item The gamma-ray spectrum is uniform and hard. The spectral index in $dN/dE\propto E^{-\gamma}$ is $\gamma \approx 1.9\pm 0.2$. The spectrum energy distribution (SED) shows a cut-off at $\sim 100$ GeV in high energy band, and a significant cut-off below 1 GeV (\citealt{Su10}), although the latter is found to be
    {less obvious in a more recent study (\citealt{Ackermann2014}).}
\item The total gamma-ray luminosity of both bubbles is $4.4^{+2.4}_{-0.9} \times 10^{37}$ erg s$^{-1}$ in 0.1--500 GeV band.
\end{itemize}

Several models have been proposed to explain the formation of the \emph{Fermi} bubbles. They usually invoke the interaction between the winds or jet and the interstellar medium to explain the morphology of the bubbles. These include: 1) winds radiatively driven from a quasar phase accretion disk in Sgr A* (\citealt{Zubovas2011}; \citealt{Zubovas2012}); 2) winds magnetically driven from a hot accretion flow in Sgr A* (\citealt{Mou2014}, hereafter paper I); 3) a jet launched from an accretion flow (\citealt{Guo1,Guo2}; \citealt{Yang2012,Yang2013}); 4) star formation winds in the Galactic center (\citealt{Crocker2011}; \citealt{Crocker2012}; \citealt{Crocker2014a, Crocker2014b}; \citealt{Carretti2013}); 5) winds produced by the periodic star capture processes in Sgr A* (\citealt{Cheng2011,Cheng2014,Cheng2015}).

In terms of the origin of gamma-ray photons of the \emph{Fermi} bubbles, theoretical models can be divided into two categories: the ``leptonic''  and ``hadronic'' ones. In the ``leptonic'' scenario, the gamma-ray photons come from the inverse Compton scattering (IC) of soft photons (interstellar radiation field and cosmic microwave background) by relativistic electrons, which often called cosmic ray electrons (CRe). The origin of CRe is different in different models. They can come from: 1)\emph{Fermi} first-order acceleration in shock front formed in the periodic star capture processes by Sgr A* (\citealt{Cheng2011,Cheng2014,Cheng2015}); 2) the \emph{Fermi} second-order acceleration through stochastic scattering by plasma instabilities (\citealt{Mertsch2011}); 3) directly carried by the jet (\citealt{Guo1}; \citealt{Guo2}; \citealt{Yang2012,Yang2013}; \citealt{Barkov2014}) or winds driven by the past star formation (\citealt{Carretti2013}).
For the leptonic scenario, the cooling timescale of CRe with energy of a few hundred GeV is only around 1 million years (\citealt{Su10}). Hence, if the CRe come from galactic center (GC), i.e., carried by jet or wind from the accretion flow of Sgr A*, the age of the \emph{Fermi} bubbles would be constrained to be less than 1-2 million years (\citealt{Guo1}; \citealt{Guo2}; \citealt{Yang2012,Yang2013}). Consequently, the power of the jet or wind must be relatively high so as to easily push the ISM away.
 This results in a high temperature of a few keV in the shocked ISM. Such a temperature is however higher than that detected in recent X-ray observations (\citealt{Kataoka2013}; \citealt{Tahara2015}).

In the ``hadronic'' model, the gamma-ray photons are produced by the decay of neutral pions produced by the inelastic collisions between cosmic ray protons (CRp) and thermal nuclei in the ISM ($pp$ collisions). The cooling timescale of CRp through $pp$ collisions is $t_{pp}\ga 3.5\times 10^{9}~{\rm yr}~(0.01~{\rm cm^{-3}} n^{-1}_{H})$. Hence this model does not require a very high kinetic power from the GC. Another advantage is that, this model can naturally explain the drop of gamma-ray spectrum close to 1 GeV (\citealt{Crocker2011}).

There are many possible origins for the CRp. They may be: 1) injected in the wind (outflows) driven by the star formation in the GC, and have accumulated for an extremely long time (e.g., a few $10^8-10^9$ years; \citealt{Crocker2011,Crocker2012,Crocker2014a,Crocker2014b}).
2) accelerated in the accretion flow by processes like weak shocks and magnetic reconnection, and then carried into the bubbles by winds of accretion flow; 3) accelerated in and injected with the jet; 4) accelerated by the multiple shocks or turbulence inside the \emph{Fermi} bubbles, or the strong shock associated with the expansion of the bubbles (e.g., \citealt{Fujita2013,Fujita2014}); 5) some diffusive injection of Galactic CRp and injection of CRp due to the expansion of the bubbles in the background ISM which is already full of CRp (\citealt{Thoudam2013}).

In paper I,  we mainly focus on how to form the bubbles, i.e., interpreting their morphology. By performing hydrodynamic simulations, we have shown that the bubbles can be successfully produced by the interaction between the winds launched from the hot accretion flow around Sgr A* and the ISM. The parameters of winds adopted in the model are taken from small scale MHD numerical simulations of accretion flows of Yuan et al. (2012). The required wind kinetic power, which is the main parameter of the model,  is also in good consistency with that obtained from independent observational constrains (see the review in \citealt{Totani2006}). In addition, the model has also successfully explained  the recent observational results of {\it ROSAT} X-ray structure (\citealt{Kataoka2013}). Most recently, the high-resolution X-ray observation to the absorption line, combined with the ultraviolet observations, obtain the bulk motion velocity and temperature of the gas around the edge of the {\it Fermi} bubbles (\citealt{FangJiang2014}). Both of them are consistent with our simulation result.

In the present work, we investigate the gamma-ray radiation of the {\it Fermi} bubbles. We adopt the ``hadronic'' scenario, i.e., the gamma-ray radiation comes from the $pp$ collisions.  For the origin of CRp, we consider the second mechanism mentioned above, i.e., CR protons are accelerated in the accretion flow (including the corona) and injected into the bubbles together with the winds. CR electrons must also be produced together with CRp.  We will include their contribution when calculating the gamma-ray radiation. However, since the energy density of CRe is much lower than CRp, we can neglect CRe in numerical simulations (see \S\ref{GammaRaySED}). The main task is then to calculate the distribution of CRp.
The motion of each CR particle is controlled by the magnetic field, so we should include magnetic field in our simulation. On the one hand, due to the scattering of magnetic irregularities in the winds, CRs are trapped in the plasma, i.e., they are ``advected'' by the wind gas. On the other hand, CRp can also diffuse with respect to the gas as they scatter off magnetic inhomogeneities. This will have an important effect on the CR distribution thus should be included in our simulation. In fact, as shown by \citet{Yang2012} and our present work, the anisotropic diffusion due to the configuration of the magnetic field lines is crucial to the interpretation of the sharpness of the observed surface brightness of the bubbles. To simulate the distribution of CRs together with the wind plasma, following \citet{Yang2012}, we adopt a simplified approach, i.e., we treat CRs as a second fluid and solve directly for the evolution of CR pressure.

The rest of the paper is organized as follows. In \S2 we briefly review our accretion wind model. In \S3, we  introduce the details of our simulation setup. The results of the simulation and the calculation of gamma-ray radiation will be presented in \S4. We then summarize and discuss our results in \S5 and \S6, respectively.

\section{Models of accretion wind and Cosmic rays}

\subsection{The Past Activity of Sgr A* and the Accretion Flow}
 Sgr A* is extremely dim currently. However, many observational evidences have been accumulated and  shown that Sgr A* very likely was much more active during the past millions of years (see reviews by \citealt{Totani2006} and \citealt{Bland-Hawthorn2013}). The mass accretion rate was estimated to be $\sim$4 orders of magnitude higher than the present value (\citealt{Totani2006}). The current mass accretion rate of Sgr A* at the black hole horizon is $\dot{M}_{\rm BH}\sim 10^{-6}\dot{M}_{\rm Edd}$, here $\dot{M}_{\rm Edd}\equiv 10L_{\rm Edd}/c^2$ is the Eddington accretion rate (\citealt{Yuan2003}; see \citealt{YuanNarayan2014} for a review of accretion models of Sgr A*). Thus the past accretion rate should be $\sim 10^{-2}\dot{M}_{\rm Edd}$.
 This rate is close to the largest accretion rate of a hot accretion flow, which is $\dot{M}_{\rm crit}\approx 0.07\alpha\dot{M}_{\rm Edd}\approx 2\times 10^{-2}\dot{M}_{\rm Edd}$ if $\alpha=0.3$ (Equation (27) in \citet{YuanNarayan2014}).
 This indicates that Sgr A* was also in a hot accretion mode in the past. Specifically, for this accretion rate, the accretion flow consists of a thin disk outside of a transition radius $R_{\rm tr}$ and a hot accretion flow inside of $R_{\rm tr}$. The value of $R_{\rm tr}$ is a function of accretion rate. For $\dot{M}_{\rm BH}\sim 10^{-2}\dot{M}_{\rm Edd}$, a reasonable value of $R_{\rm tr}$ is $\sim 200R_s$ (\citealt{YuanNarayan2004}). This is the value we adopt in our fiducial model.

\subsection{Winds from Hot Accretion Flows}

 One of the most important findings in the theoretical study of hot accretion flow in the past few years is that strong winds exist (\citealt{YuanBW2012a}; \citealt{Narayan2012}; \citealt{LiOS2013}; \citealt{Stone99}; \citealt{BlandfordBegelman1999}; \citealt{Begelman2012};  \citealt{Sadowski2013}; see review by \citealt{Yuan2015}). This has been confirmed by radio polarization (\citealt{Aitken2000}; \citealt{Bower2003}; \citealt{Marrone2007}) and X-ray observations (\citealt{Wang2013}). For example, it is found that only about $1\%$ of the gas available at the Bondi radius finally falls onto the black hole while the rest is lost via winds. The detailed properties of winds, such as the velocity, mass flux, and angular distribution, have been obtained by magnetohydrodynamic (MHD) simulations of hot accretion flows (\citealt{Yuan2015}; \citealt{YuanBW2012a}). Significant winds are found from $20R_s$ up to the outer boundary of the hot accretion flow, which is $R_{\rm tr}$ in the present case. The mass flux of winds is described by (\citealt{Yuan2015}):
 \begin{equation}
  \dot{M}_{\rm wind}(R)=\dot{M}_{\rm BH}\frac{R}{20R_s}.
 \end{equation}
 For $R_{\rm tr}\sim 200R_s$ and $\dot{M}_{\rm BH}\sim 0.02\dot{M}_{\rm Edd}$,
  we have $\dot{M}_{\rm wind}\sim 20\%\dot{M}_{\rm Edd}$. We set $\dot{M}_{\rm wind}$ to be $18\%\dot{M}_{\rm Edd}$ in our simulations.
 The terminal poloidal velocity of winds is determined by the radius where the wind is launched (\citealt{Yuan2015}),
 \begin{equation}
  v_{\rm winds}(R)\approx (0.2-0.4)v_k(R).
 \end{equation}
 Here $v_k(R)=(GM/R)^{1/2}$ is the Keplerian velocity at radius $R$. At any given radius, say $R_0$, the poloidal velocity of winds is a mixture of winds launched from the region of $r\la R_0$, and these winds have different poloidal velocities described by Equation (2). In the present work, we simply assume that the poloidal velocity of winds is simply described by $\sim v_k(R_{\rm tr})\sim 4.6\%c$ since most of the mass flux of winds originate from around $R_{\rm tr}$.
 Thus the kinetic power of the thermal gas in the winds is $10^{42} {\rm erg~s^{-1}}$. The internal energy is negligible compared with the kinetic energy for the winds.

\subsection{Production of Cosmic Rays in the Accretion Flow}
\label{crinaccretion}
The detailed acceleration process of cosmic rays (CRs) is beyond the scope of the present work.
Here we assume that CRs are injected from the origin, together with the winds launched from the accretion flow of Sgr A*. These CRs may have the following origins. First, within the accretion flow the magnetic field is tangled, and the accretion flow is turbulent. Physical processes within the accretion flow such as MHD turbulence, magnetic reconnection, and also weak shocks can accelerate some particles into relativistic energies. In fact, numerical simulations have shown the existence of magnetic reconnection in the hot accretion flows (e.g., \citealt{Machida2003}) and  acceleration of particles has been studied (e.g., \citealt{Ding2010}).
Second,  at the coronal region of the accretion flow, magnetic reconnection occurs perhaps more frequently. Loops of magnetic filed emerge into the corona from the accretion flow due to Parker instability. Since their foot points are anchored in the accretion flow which is differentially rotating and turbulence, reconnection occurs subsequently, as have been shown both analytically and numerically (\citealt{Romanova1998}; \citealt{Blandford2002}; \citealt{Hirose2004}; \citealt{Goodman2008}; \citealt{Uzdensky2008}; \citealt{Yuan2009}). These events will efficiently accelerate particles and produce CRs. Because winds are  launched at the surface of the accretion flow (\citealt{Yuan2015}), we expect that  many CRs must be contained in the winds.
The ratio of the power of CRs (mainly CRp; the power of CRe can be neglected) and the total power of the winds (thermal gas and CRs) is defined as $\eta_{\rm CR}$ --- the ``CR parameter''.
Since the details of such particle acceleration are still poorly understood, in the present work we treat $\eta_{\rm CR}$ as a free parameter.

\section{Numerical SIMULATION}

 \subsection{Equations}
 The MHD equations describing CR advection, diffusion, and dynamical coupling between the thermal gas and CRp are as follows,
\begin{gather}
  \frac{d \rho}{d t} + \rho \nabla \cdot {\bf v} = 0,\label{hydro1} \\
  \rho \frac{d {\bf v}}{d t} = -\nabla (P_{1}+P_{2}+P_{B}) -\rho \nabla \Phi +\nabla \cdot {\bf T},\label{hydro2}
 \end{gather}
 \begin{gather}
  \frac{\partial e_{1}}{\partial t} +\nabla \cdot(e_{1}{\bf v})=-P_{1}\nabla \cdot {\bf v}+{\bf T}:\nabla {\bf v}+\mathcal{L}_{c}, \label{hydro3} \\
  \frac{\partial e_{2}}{\partial t}+\nabla \cdot(e_{2}{\bf v}-{\bf \kappa} \cdot \nabla e_{2}) =-P_{2}\nabla \cdot {\bf v}, \\
  \frac{\partial \bf B}{\partial t}-\nabla \times \bf (v \times \bf B) =0, \\
 {\bf T}=\mu [\nabla {\bf v}+(\nabla {\bf v})^{tr}-\frac{2}{3} {\bf I} \nabla \cdot {\bf v}].
 \end{gather}
\\
 Here $P_{1}=(\gamma_{1}-1)e_{1}$ and $P_{2}=(\gamma_{2}-1)e_{2}$ are thermal pressure and CRp pressure respectively, in which $\gamma_{1}=5/3$, $\gamma_{2}=4/3$ are adiabatic indices for ideal gas and CRp. $P_{B}=B^{2}/8\pi$ is the magnetic pressure and $\mathcal{L}_{c}$ is cooling function.  In order to avoid the density near the GC being too high, the gravitational force is adopted in the form of
 \be \nabla \Phi=-\frac{2\sigma^{2}}{r+r_{0}}\vec{r}, \ee
 where $r_{0}$ is set to be 0.1 kpc.

The diffusion coefficient tensor $\kappa$ can be written as:
 \be
 \kappa_{ij}=\kappa_{\perp}\delta_{ij}-(\kappa_{\perp}-\kappa_{\parallel})\frac{B_{i}B_{j}}{B^2},
 \ee
 where $\kappa_{\perp}$ and $\kappa_{\parallel}$ are the diffusion coefficients perpendicular and parallel to the magnetic field, respectively. The relationship between the two coefficients is given by
 \be
 \frac{\kappa_{\perp}}{\kappa_{\parallel}}=\frac{1}{1+(\lambda_{\parallel}/r_{g})^2}, \label{K_perp}
 \ee in which $\lambda_{\parallel}$ and $r_{g}$ are the parallel mean free path and the particle Larmor radius. In our case, $\lambda_{\parallel}$ is much larger than $r_{g}$, hence, $\kappa_{\perp} \ll \kappa_{\parallel}$. Therefore we set $\kappa_{\perp}$ to be 0 in most of our simulations. We should note that some works have shown that even if the turbulence level $\eta = \langle\textbf{B}^{2}\rangle/(B^{2}_{0}+\langle\textbf{B}^{2}\rangle)$ ($\textbf{B}$ is a random magnetic field) is small, say 0.1, the transverse diffusion coefficient $\kappa_{\perp}$ is much more effective than the quasi-linear result described  by Equation (\ref{K_perp}), especially when the maximum scale of turbulence is comparable to the Larmor radius (e.g., \citealt{Casse2001}). In this case, the transverse diffusion can not be neglected. Considering the complexity of magnetic field coexisting with the CRs on the microscope scale, in one test we also include a non-zero transverse diffusion (refer to Table \ref{table1} for details).

 We include the cooling effect. We assume that the abundance of the halo is [Fe/H] = $-0.5$ (\citealt{Miller2013}). The cooling function $\mathcal{L}_{c}$ is set to be the cooling curve in collisional ionization equilibrium state (\citealt{Sutherland1993}). Cooling effect may be strong for central molecular zone (CMZ; \citealt{Morris1996}, see paper I for its setup) gas, since the density is very high there. But the heating effects for CMZ gas may also be strong, say radiative heating from stars and shock heating from supernovae,  which are not considered here. Here for simplicity, we assume that CMZ gas does not suffer from cooling and heating effects. We set a ground temperature of $10^{4}$ K, below which cooling disappears.

 The viscosity is also included. As we have shown in paper I, if viscosity were not included in our model, both the Rayleigh-Taylor (RT) and Kelvin-Helmholtz (KH) instabilities would have developed  into large-scale structures during the formation of the \emph{Fermi} bubbles. Viscosity may be affected by the magnetic field. Specifically, viscosity perpendicular to the magnetic field is strongly suppressed. However, we do not consider this complexity and still set an isotropic viscosity as in paper I and \citet{Guo2}. The dynamical viscosity coefficient $\mu$ is adopted to be 1.7 $\muunit$, which is slightly different from the value of 2 $\muunit$ in the basic run in paper I.

 We neglect the influence of CRs on the magnetic field, which is complicated and difficult to be handled  in our code.

 \subsection{Simulation Setup}

 We use the ZEUS3D code (\citealt{Clarke1996, Clarke2010}, also see  http://www.ica.smu.ca/zeus3d/), to solve the above equations. ZEUS3D code is an Eulerian computational fluid dynamics code written in FORTRAN, and is a member of the ZEUS-code family. It can be used to deal with two-fluid problem. It adopts finite-difference method on a staggered grid with an accurate of the second order, and uses time-explicit, operator splitting solution in which the solution procedure is grouped into source step and transport step (see \citealt{Stone92} for details). For the diffusion process, we directly difference the diffusion term in the source step.
 We adopt 3-D Cartesian coordinates. Computational domain is almost the same with paper I:
  $-5.1~{\rm kpc} \sim +5.1~{\rm kpc}$ in $X$-, $Y$-directions, and 0 kpc $\sim$ +11.7 kpc in $Z$-direction, and the $X-Y$ plane is the Galactic plane while the $Z$-axis stretches along the Galactic pole. Sgr A* is just located in the origin. We adopt non-uniform grid, with $\bigtriangleup x_{i+1}/\bigtriangleup x_{i}=1.056$ for $x>0$, $\bigtriangleup x_{i+1}/\bigtriangleup x_{i}=0.9470$ for $x \leq 0$ , $\bigtriangleup y_{j+1}/\bigtriangleup y_{j}=1.056$ for $y>0$  $\bigtriangleup y_{j+1}/\bigtriangleup y_{j}=0.9470$ for $y \leq 0$, and $\bigtriangleup z_{k+1}/\bigtriangleup z_{k}=1.0285$.
  The numbers of meshes are 120, 120 and 119 in $X$-, $Y$-, $Z$-direction, respectively. Since most of the parameters are the same with paper I, in the following subsections we mainly introduce the different parts.

  The global time step in the simulation is determined by the CFL condition (\citealt{Clarke1996}). When viscosity, diffusion and cooling processes are considered, the time step is calculated by:
  \be
  {\rm dt=min\{dt(CFL), dtvisc, dtdf, dtcooling\}  }
  \ee
  in which the time step of viscosity process---dtvisc is calculated by $\rho (\Delta x)^{2}/\mu$ (also see \citealt{Guo2}), diffusion process---dtdf is calculated by Cdf min\{$dx_{i}dx_{j} B^{2}/(\kappa_{\parallel} B_{i}B_{j}$)\} where Cdf is chosen to be 0.25, and the cooling time step---dtcooling is neglected since it is much larger than the others. 

 \subsection{Initial Conditions}

 The initial density distribution is set in the form of $n_{e}=n_{e0}/r^{\alpha}_{{\rm kpc}}$ ($r_{\rm kpc}=r$/1 kpc), where $\alpha$ is set to be 1.6. The values of $n_{e0}$ is set to be $7\times 10^{-2}~{\rm cm^{-3}}$ except run C (see Table \ref{table1}). The initial density is significantly higher than that in paper I, but close to the most recent work of \citet{Miller2015}.
 Besides, we set the gradient of the total pressure (thermal pressure plus magnetic pressure) to balance the gravitational force.

 \be
 P_{tot}({\vec r})=P_{1}+P_{B}=1.25 \sigma^{2} \rho({\vec r})
 \ee
 where $P_{tot}$, $P_{1}$ and $P_{B}$ (see section 3.4) are the total pressure, thermal pressure and magnetic pressure of the initial ISM.
 Therefore the initial ISM is not isothermal. We have also included the massive CMZ in our initial conditions (see paper I for the setup of CMZ), 
  which plays an important role in collimating the winds. CRs are not included in the initial ISM, but injected from the GC (see Section 3.5).

\begin{table*}
  \centering
  \begin{minipage}{130mm}
  \renewcommand{\thefootnote}{\thempfootnote}
  \caption{{\rm  Simulations Parameters and Results}}
  \centering
  \begin{tabular}{@{} c  c  c  c  c  c  c  c  c  c}
  \hline
        & $n_{e0}$
        & {$v_{wind}$ \footnote{the velocity of winds.}}
        & {$\dot{M}_{wind}$ \footnote{the mass outflow rate of winds.}}
        & $\eta_{CR}$ \footnote{the ratio of the power of CR to the total power of winds (thermal gas + CR).}  
        & $\kappa_{\perp}/\kappa_{\parallel}$
        & {$t_{\rm FB}$ \footnote{the age of the \emph{Fermi} bubbles.}}
        & {$I_{\gamma}$ \footnote{ the gamma-ray intensity averaged on the projected map and integrated from 0.1GeV to infinity. We have included the contribution from the secondary and primary leptons.}}
        & {$L_{\gamma}$ \footnote{the simulated gamma-ray luminosity of the \emph{Fermi} bubbles at $E_{\gamma}\geq$ 0.1 GeV. We have included the IC process as in $I_{\gamma}$. A fixed solid angle of 0.70 sr for both bubbles is assumed according to  \citet{Ackermann2014}. The center of each bubble is set to be $(R,~|z|)=(0,~5~{\rm kpc})$.}}
        \\
    Run & (${\rm cm^{-3}}$) &  (km/s) & ($\medd$) &  &  & ({\rm Myr}) & (${\rm GeV~cm^{-2}~s^{-1}~sr^{-1}}$) & (${\rm erg/s}$) &   \\
    \hline
     A & $7.3\times 10^{-2}$ & $1.4\times 10^{4}$ & 18\% & 50\% & 0   & 7.0 & $3.3\times 10^{-6}$ & $4.4\times10^{37}$  \\ 
     B & $7.3\times 10^{-2}$ & $1.4\times 10^{4}$ & 18\% & 50\% & 0.01& 7.6 & $3.8\times 10^{-6}$ & $5.0\times10^{37}$  \\ 
     C & $3.7\times 10^{-2}$ & $1.4\times 10^{4}$ & 18\% & 50\% & 0   & 6.3 & $1.3\times 10^{-6}$ & $1.7\times10^{37}$  \\ 
     D & $7.3\times 10^{-2}$ & $1.4\times 10^{4}$ & 18\% & 33\% & 0   & 7.9 & $1.6\times 10^{-6}$ & $2.1\times10^{37}$  \\ 
     E & $7.3\times 10^{-2}$ & $7.0\times 10^{3}$ & 73\% & 50\% & 0   & 7.6 & $3.9\times 10^{-6}$ & $5.1\times10^{37}$  \\ 
    \hline
 \label{table1}
  \end{tabular}
 \end{minipage}
\end{table*}

 \subsection{Magnetic Field}

We assume that the configuration of the magnetic field in GC in the past is the same with the present time. We also assume for simplicity that the winds launched from the accretion flow do not contain magnetic field. For the initial galactic magnetic field (GMF), we refer to the work of \citet{Jansson2012} for details. In our simulations, we assume the initial GMF contains two components, i.e., a large-scale regular field and a small-scale random field.

 The large-scale regular field can be divided into three components: a disk component, a toroidal halo component, and an out-of-plane component. The disk component is concentrated in the galactic disk, and drops into very weak magnetic field at height larger than 0.7 kpc. Hence the disk component has a weak influence on the \emph{Fermi} bubbles, while the toroidal and the out-of-plane component in the halo may play an important role. For simplicity, in our simulations, we only consider these two large-scale regular halo components, and superpose a tangled component as a small-scale random field:
 \begin{gather}
  \vec{B}=\vec B_{tor}+\vec B_{X}+\vec B_{tb}. \label{Btot}
 \end{gather}
 where $\vec B_{tor}$ is the toroidal halo component, $\vec B_{X}$ is the out-of-plane component, and $\vec B_{tb}$ is the tangled field.
 The details of the three components are described in the Appendix.

\subsection{Cosmic Ray}

As we state in \S2.3, we use a parameter $\eta_{\rm CR}$ to describe the energy fraction of the CRp in the wind. We typically adopt $\eta_{\rm CR}=50\%$ in our simulations. But one lower value is also chosen in one model (see Table 1).

The diffusion coefficient of CRs is energy dependent. A typical value by fitting CR data is $\kappa_{iso}=(3-5)\times 10^{28}~{\rm cm^{2}~s^{-1}}$ for $\sim$1 GeV CR particles in a sufficiently tangled magnetic field on small scales, and scales as $E_{cr}^{0.3}\sim E_{cr}^{0.6}$, with $E_{cr}$ being the energy of a CR particle (\citealt{Strong2007}). We simply treat the diffusion coefficients as independent of energy in our simulations. As argued in \citet{Yang2012}, the  diffusion coefficient $\kappa_{iso}$ in a tangled magnetic field may be suppressed compared with $\kappa_{\parallel}$ by a factor of 3. We therefore set $\kappa_{\parallel}$ to be $1\times10^{29}~\rm{cm^{2}~s^{-1}}$ here, which is suitable for CRs at energies of a few GeV.

\section{RESULTS}
\subsection{Distribution of CR protons}

 \begin{figure*}[!htb]
 \centering
 \begin{center}
 \includegraphics[width=0.45\textwidth]{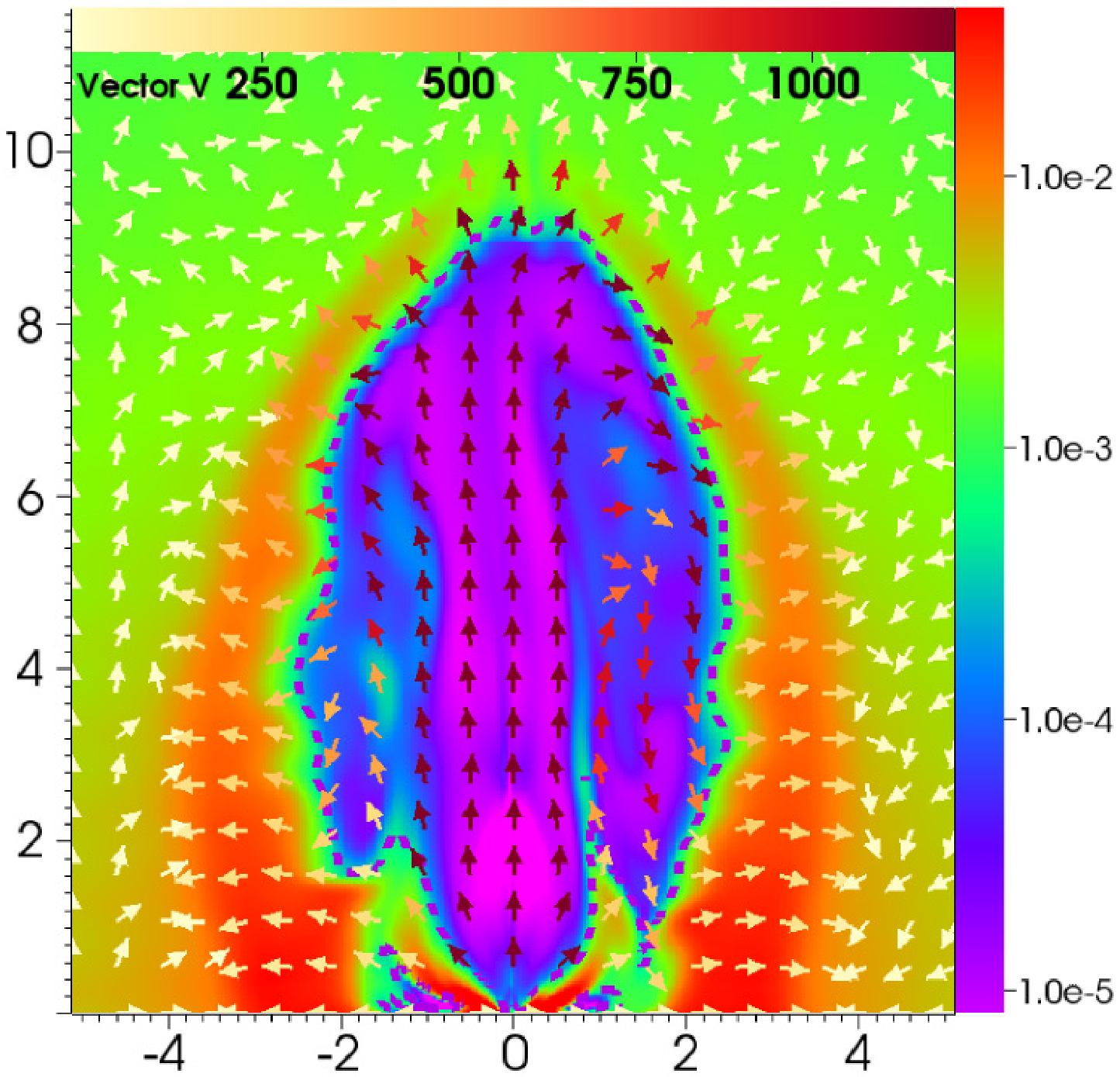}
 \includegraphics[width=0.45\textwidth]{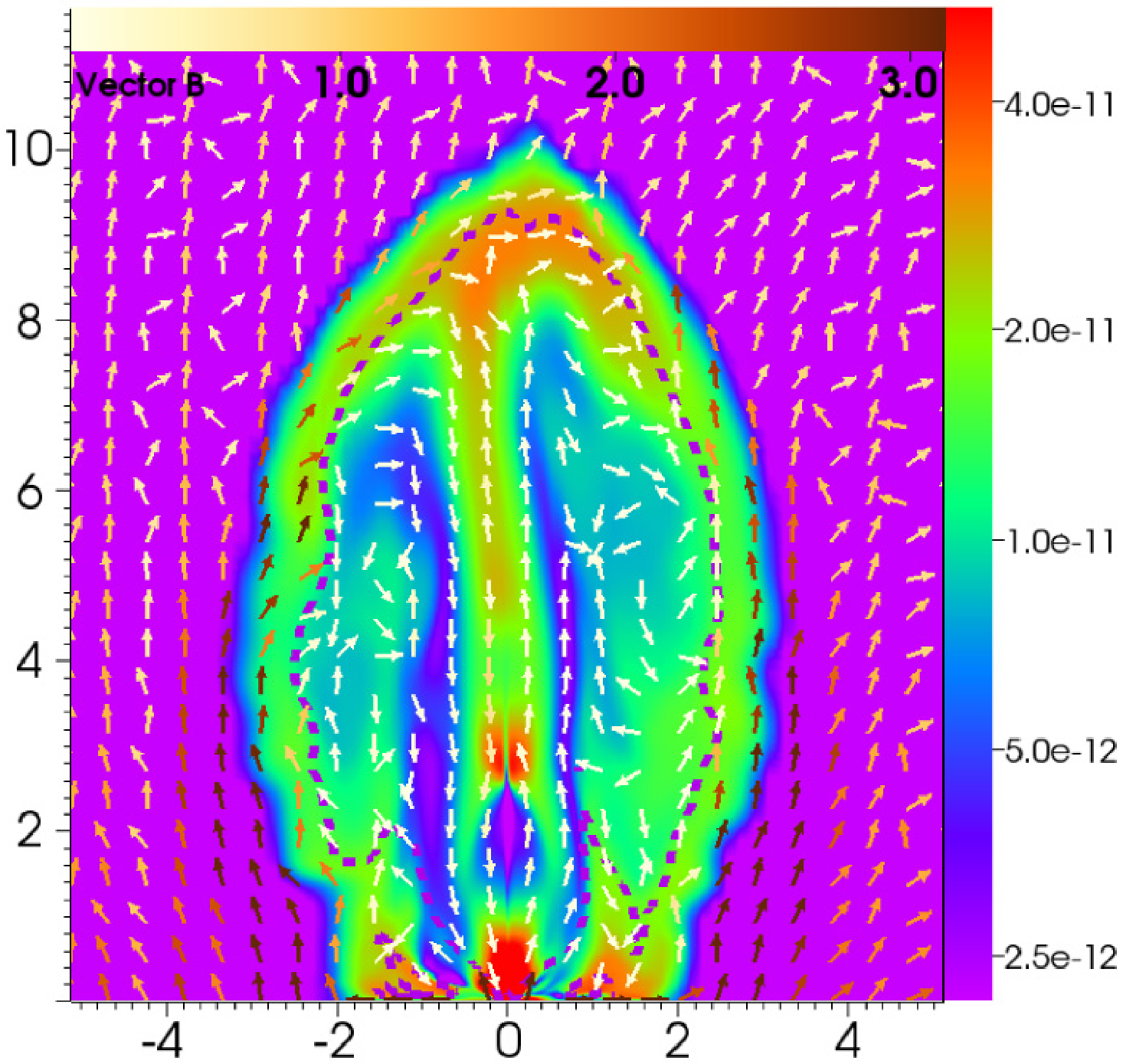}
 \includegraphics[width=0.45\textwidth]{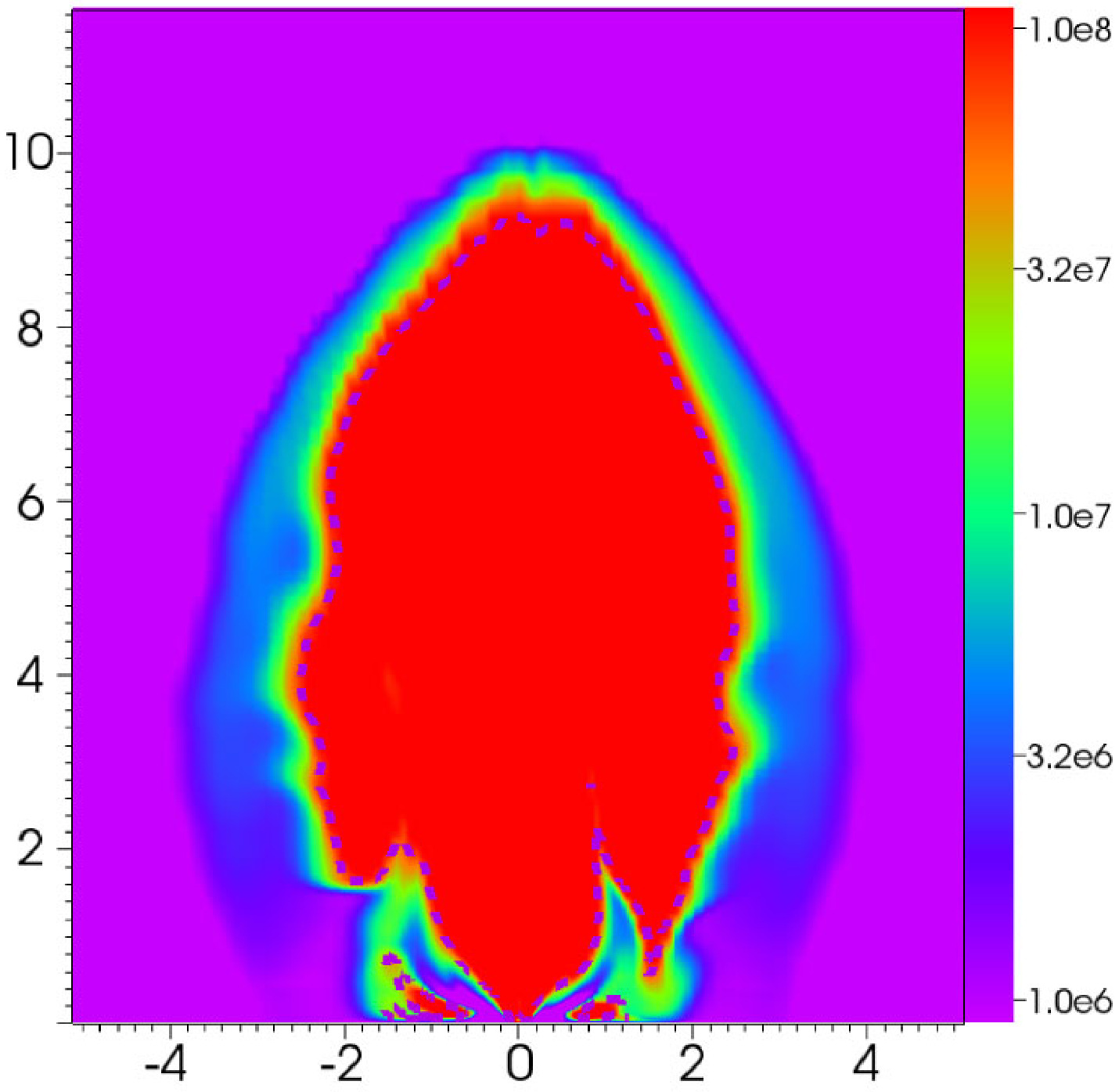} \\
 \vspace{-0.6cm}
 \end{center}
 \caption{$X-Z$ sectional views of the results of run A at $t=7$ Myr. Coordinates are in units of kpc. \emph{Left}: number density distribution of thermal electrons ($n_e$) in units of cm$^{-3}$. Velocity vectors are also plotted, with the color bar at the top of the plot denoting the value of velocity in units of ${\rm km~s^{-1}}$. \emph{Middle}: energy density distribution of CRp (e2) in units of $\rm erg~cm^{-3}$. Magnetic field vectors are also plotted in this map in units of $\mu$G. \emph{Right}: temperature in units of Kelvin. The dashed lines in these three maps denotes the contact discontinuity (CD).  }
 \label{plot1}
\end{figure*}

\begin{figure}[!htb]
  \centering
  \includegraphics[width=0.49\textwidth]{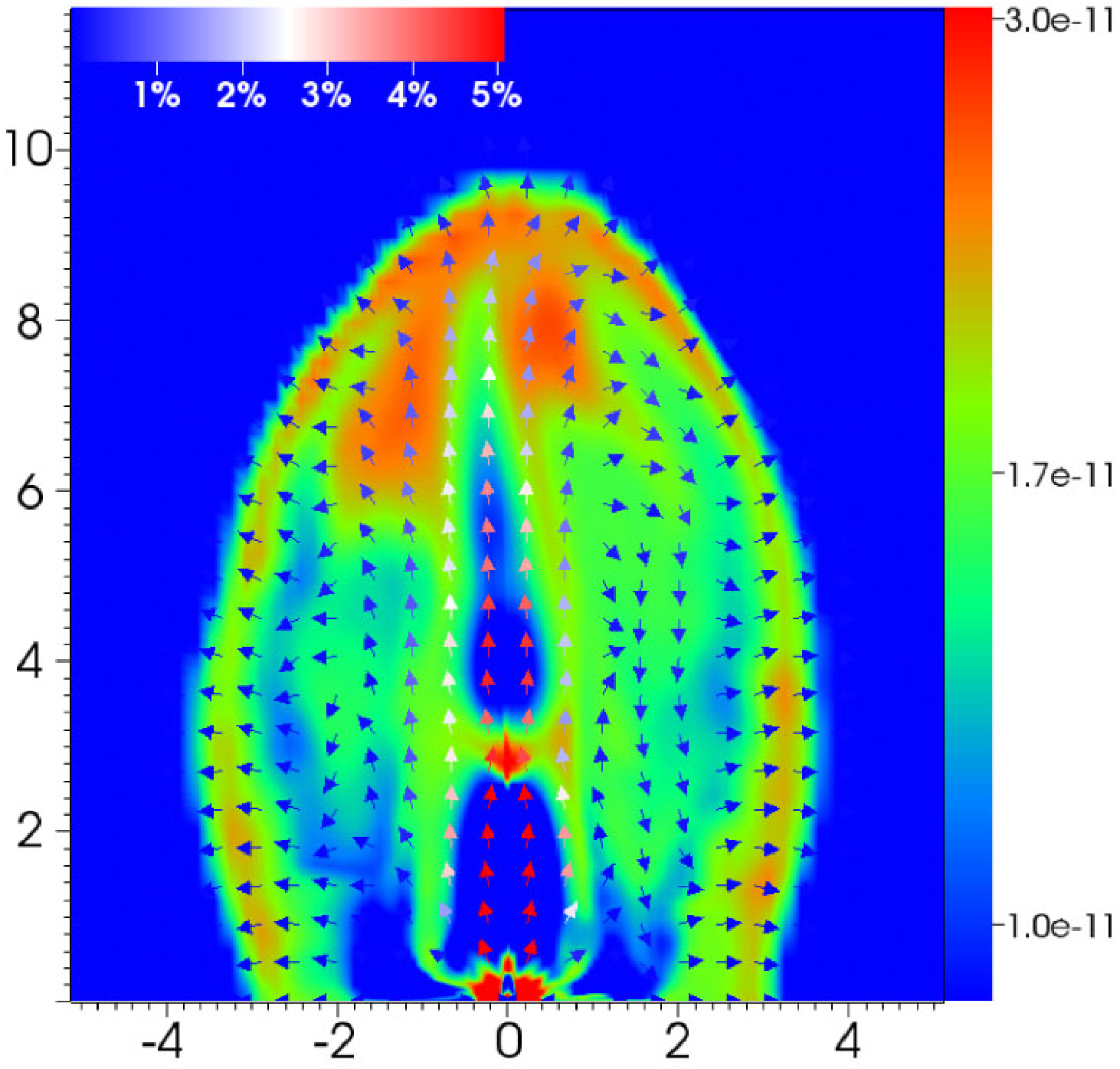}
 \vspace{-0.6cm}
 \caption{Distribution of thermal energy density of run A at $t=7$ Myr. The values in the vertical color bar are in units of erg~cm$^{-3}$. The horizontal color bar is for the velocity, which is denoted by arrows, and its value is in units of speed of light. }
 \label{plot2}
\end{figure}

 The parameters of various models we have simulated are given in Table 1. We choose run A as our fiducial model, in which the age of 7 Myr is close to paper I. Figure \ref{plot1} shows the distributions of density of thermal gas ($n_{e}$), energy density of CRp, and the temperature on $X-Z$ plane. The density distribution here is similar to paper I. The density of the shocked ISM behind the forward shock is enhanced, and the density sharply decreases crossing the contact discontinuity (CD; denoted by the \emph{purple} dashed line in Figure \ref{plot1}) towards the interior of the bubbles. Inside the CD, the bubble is filled with gas with very low density and very high temperature (a few $10^{8}$--$10^{9}$ K).
 The CRp pressure decreases away from the CD towards the exterior of the bubbles. Near the CD, the CRp pressure is comparable to the thermal pressure of the shocked ISM, and it expels the thermal gas away from the CD, leaving a zone with density somewhat lower than the ``typical'' density of the shocked ISM. We call it a ``permeated zone''. Therefore the shock structure in the present two-fluid MHD case is slightly different from that in paper I (see Figure 1 in paper I).
 The temperature in the shocked ISM ranges from $3\times 10^{6}$ K to $1\times 10^{7}$ K when going from $z=3$ kpc to 9 kpc. Since the radiation from the shocked ISM dominates the observed X-ray flux, we thus expect that observed X-ray spectrum should be fitted by the radiation from a multi-temperature medium. This seems to be consistent with the most recent {\it Suzaku} X-ray observation (\citealt{Tahara2015}).

 The magnetic field in the shocked ISM is roughly parallel to the CD, which agrees with the results of \citet{Yang2012}. However, the alignment is not perfect, which is crucial for the hadronic scenario, as we will discuss later. Under such kind of magnetic field configuration, CRs can not diffuse too far away from the CD. Therefore the morphology of the bubbles is determined by the CD and this is also the reason why the edges of the \emph{Fermi} bubbles look sharp.

 As shown by the middle plot of Figure 1, the distribution of CRp looks like an ``umbrella-like'' structure. At the beginning, CRp move in the direction of the polar axis, and form the ``handle'' of the ``umbrella''. After reaching the top, CRp fall down along the CD. This is because, CRp can only diffuse along the magnetic field lines which is roughly parallel to the CD. The density of CRp in the large volume enclosed by the umbrella-like structure is quite low, while the thermal pressure in this region is relatively high compared to the CR pressure (see Figure \ref{plot2}). This is of course because of the total pressure balance. A question is,  the winds are blown out isotropically, why do CRp move like a ``jet'' at the beginning? Combining the middle panel in Figure \ref{plot1} and Figure \ref{plot2}, we find that, the CRp ``handle'' along the polar axis is surrounded by a ``wall'' of high thermal pressure.
 The relatively high thermal pressure arises because of the strong viscous heating at the interface between the high-speed winds and the  blown-up CMZ gas.
 In other words, the viscously heated high-pressure gas pushes the winds and form the ``handle''. Besides, the magnetic field inside the winds (the \emph{purple} region in the left panel of Figure \ref{plot1}) is parallel to the movement direction of the winds. This prevents the diffusion of CRs in the winds in the horizontal direction.

We find that in the ``umbrella face'' the density of CRp decreases from the high to low latitudes, while it is on the opposite for the density of thermal gas. Our calculation shows that such a distribution can generate a roughly uniform gamma-ray emissivity shell in the sectional map thus a projection map with a constant brightness along the latitude (see \S 4.3 for details). If on the other hand the density of CRp did not increase with the latitude, the gamma-ray emissivity would have significantly decreased with the height. This is because in either the hadronic or leptonic scenario, the density of both thermal gas and the radiation field decreases with the latitude, thus the surface brightness would have become dimmer with the increasing latitude. One advantage of this kind of ``umbrella handle'' structure is that it can keep the total energy of CRp from serious loss due to sideways adiabatic expansion. If on the other hand the CRp would follow a free and adiabatic expansion, more energy of CRp would have been lost, and it would be difficult to generate enough gamma-ray photons.

As we have stated before, the magnetic field is not strictly parallel to the CD (see Figure \ref{plot1}). This allows some CRp to diffuse into the shocked ISM, form a permeated zone outside the CD. We find that, about 60\% of the total CRp have diffused outside the CD into the permeated zone, and the remaining 40\% are still constrained inside the CD. This is another crucial point for the hadronic scenario. The permeated zone is the dominant region to generate the gamma-ray emission because of the relatively high density of thermal nuclei there, and only the CRp in the permeated zone are the ``effective'' energy source for the gamma-rays.

 \subsection{Gamma-ray Radiation}
 \label{GammaRaySED}

 As we know, if the energy of the CRp is higher than a threshold value, the collisions between CRp and thermal protons can produce pions. The reaction channels are:
 \begin{gather}
 p+p \rightarrow p+p+a \pi^{0}+b(\pi^{+}+\pi^{-}), \\
 p+p \rightarrow p+n+\pi^{+}+a\pi^{0}+b(\pi^{+}+\pi^{-}),
 \end{gather}
 where $a$ is generally equal to $b$ and they denote the number of pions produced in the reaction. Each neutral pion will instantly decay into two $\gamma$-ray photons:
 \be
 \pi^{0} \rightarrow 2\gamma .
 \ee
 This process induces a lower limit of the gamma-ray photons, which is about half of the energy of a neutral pion. Since the rest mass of a pion is 135 MeV$c^{-2}$, the gamma-ray spectrum will have a cutoff near $\sim70$ MeV, which is a characteristic signature for the pion-decay, as observed in some supernova remnants (\citealt{Ackermann2013}). The gamma-ray spectrum of the \emph{Fermi} bubbles also shows such a hardening feature below $\sim 1$ GeV, as highlighted in \citet{Crocker2011}. This is a possible evidence for the hadronic scenario.
 The charged pions produced in $pp$ collisions also contribute to the gamma-ray emission by generating high-energy secondary leptons. Charged pions decay in the follow ways:
 \begin{gather}
 \pi^{+} \rightarrow \mu^{+}+\nu_{\mu},~~ \mu^{+} \rightarrow e^{+}+\nu_{e}+\bar{\nu_{\mu}}, \label{pi1} \\
 \pi^{-} \rightarrow \mu^{-}+\bar{\nu_{\mu}},~~ \mu^{-} \rightarrow e^{-}+\bar{\nu_{e}}+\nu_{\mu}. \label{pi2}
 \end{gather}
 All of these reactions can be finished instantaneously (the mean lifetime is $2.6\times 10^{-8}$ s for charged pions, and $2.2\times 10^{-6}$ s for muons). The secondary positrons and electrons are also of high energies, and they can scatter with the seed photons and produce gamma-rays.

 The products of $pp$ collisions are calculated using the cparamlib package\footnote{\url{https://github.com/niklask/cparamlib}}, and the differential cross sections in $pp$ collisions are from \citet{Kamae2006}.
  We assume that the energy distribution of CRp follows such a power-law form:
 \be
  dn_{p}(T_{p})/dT_{p}=N_{p,pl} T^{-N}_{p} ~~(1{\rm GeV}\leq T_{p}\leq 2{\rm TeV}),
 \ee
 where $dn_{p}$ is the number density of CRp with kinetic energies between $T_{p}$ and $T_{p}+dT_{p}$,  $N_{p,pl}$ is a constant, $T_{p}$ is the kinetic energy of CRp, and the index $N$ is set to be 1.90.  
 The number of final particles (including gamma-rays, secondary leptons and neutrinos) produced per unit volume, per unit time, and per unit energy in $pp$ collision is given by (\citealt{Ackermann2014}):
 \be
  \frac{dQ_{f}}{dE_{f}}=\int\frac{d\sigma(T_{p},E_{f})}{dE_{f}}n_{H}c\frac{dn_{p}}{dT_{p}}dT_{p}, \label{ppreaction}
 \ee
 where the subscript $f$ denotes the final particle species (such as $\gamma$, $e^{+}$, and $e^{-}$), $E_{f}$ is the energy of the final particle, $d\sigma(T_{p},E_{f})/dE_{f}$ is the differential inclusive cross section (\citealt{Kamae2006}), $n_{H}$ is the number density of Hydrogen nuclei, and $c$ is light speed.

 The secondary leptons undergo cooling via IC and synchrotron radiation. The evolution of the energy distribution of secondary leptons can be obtained by:
 \be
 \frac{\partial n_{s}(E_{s},t)}{\partial t}+\frac{\partial}{\partial E_{s}}[\dot{E_{s}}n_{s}(E_{s},t)]=\frac{dQ_{s}}{dE_{s}}, \label{evolution}
 \ee
 where the subscript $s$ means secondary positions or electrons, $n_{s}$ means the number density of the secondary leptons per unit energy interval, and $E_{s}$ is the energy of the secondary leptons. $dQ_{s}/dE_{s}$ is the source term (see Equation (\ref{ppreaction})). The energy lost rate $\dot{E_{s}}$ includes IC and synchrotron processes, with the magnetic field being 2.5 $\mu$G according to Figure \ref{plot1}. We choose the stationary solution of this equation to calculate the IC scattering, and the seed photons are set to be the radiation field at $(R, z)$ = (0, 5 kpc) (\citealt{PorterStrong2005}).

 As we have stated in \S2.3, CRe can also be produced at the same processes as of producing CRp; thus there must be some CRe contained in the winds. The age of the \emph{Fermi} bubbles in our model is comparable with the cooling timescale of CRe with energy of $\sim$ 30--100 GeV (see Figure 28 in \citealt{Su10}), so CRe with energy less than this value can still exist in the bubble. Based on these considerations, we can assume some primary CRe in the following power-law form:
 \be
  dn_{pe}(E)/dE =N_{pe,pl} E^{-\Gamma}e^{-E/E_{cut}},
 \ee
 where $dn_{pe}$ is the number density of the primary CRe with energies between $E$ and $E+dE$, $E$ is the energy of primary CRe, $\Gamma$ is set to be 2.1, and the cut-off energy $E_{cut}$ is set to be 60 GeV. The energy density of the primary CRe integrated from $\sim$ 1 MeV to 60 GeV is $7\times10^{-15}$ erg cm$^{-3}$, which is about 3 orders of magnitude lower than that of CRp.

 The number of gamma-ray photons produced per unit volume, time, and energy, which is the sum of $\pi^{0}$ decays and IC scattering of the leptons, is given by:
 \begin{gather}
  \frac{dN_{\gamma}}{dE_{\gamma}}=\frac{dQ_{\gamma}}{dE_{\gamma}}+{\rm IC~}(se^{+}+se^{-}+pe^{-}),  \\
 {\rm IC(e)}
 =c \int \frac{d \sigma_{IC}(E_{\gamma}, E_{e}, E_{ph})}{dE_{\gamma}} \frac{dn_{e}}{dE_{e}} dE_{e} \frac{dn_{ph}}{dE_{ph}} dE_{ph}
 \end{gather}
 where $se^{+}$, $se^{-}$ and $pe^{-}$ mean secondary positrons, secondary electrons and primary CR electrons, respectively. The cross sections of IC scattering, $d \sigma_{IC}(E_{\gamma}, E_{e}, E_{ph})/dE_{\gamma}$  is given by \citet{BlumenthalGould1970}, and we use the radiation field model at $(R, z)$ = (0, 5 kpc) to set the seed photons $dn_{ph}/dE_{ph}$ (\citealt{PorterStrong2005}). The total gamma-ray intensity, which is the gamma-ray energy observed by the telescope during per unit time, per unit energy in the vicinity of $E_{\gamma}$, per unit solid angle, per unit area sensor is then
 \be
  I_{\gamma}=\int j_{\gamma}dl=\frac{1}{4\pi}\int E_{\gamma}\frac{dN_{\gamma}}{dE_{\gamma}} dl, \label{Intensity}
 \ee
 where $j_{\gamma}$ is the emissivity per unit solid angle, $dl$ is the length element along the line-of-sight. As mentioned above, $I_{\gamma}$ contains three components: gamma-rays from $\pi^{0}$ decays $I_{\pi^{0} \rightarrow \gamma}$, IC scattering of secondary leptons $I_{se \rightarrow \gamma}$, and IC scattering of primary CR electrons $I_{pe \rightarrow \gamma}$.

\begin{figure}[!htb]
  \centering
  \plotone{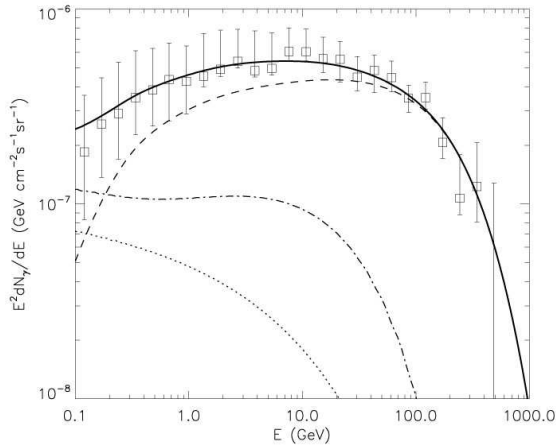}
  \caption{The gamma-ray spectral energy distribution calculated based on run A. The rectangles with error bars show the latest observational results (\citealt{Ackermann2014}). The solid line is the sum of the dashed (for the $\pi^{0}$ decays), dotted (for IC process of the secondary leptons generated in hadronic reaction), and dot-dashed (for IC of the primary electrons) lines. }
  \label{plot3}
\end{figure}

 \begin{figure*}[!htb]
 \centering
 \begin{center}
 \includegraphics[width=0.3\textwidth]{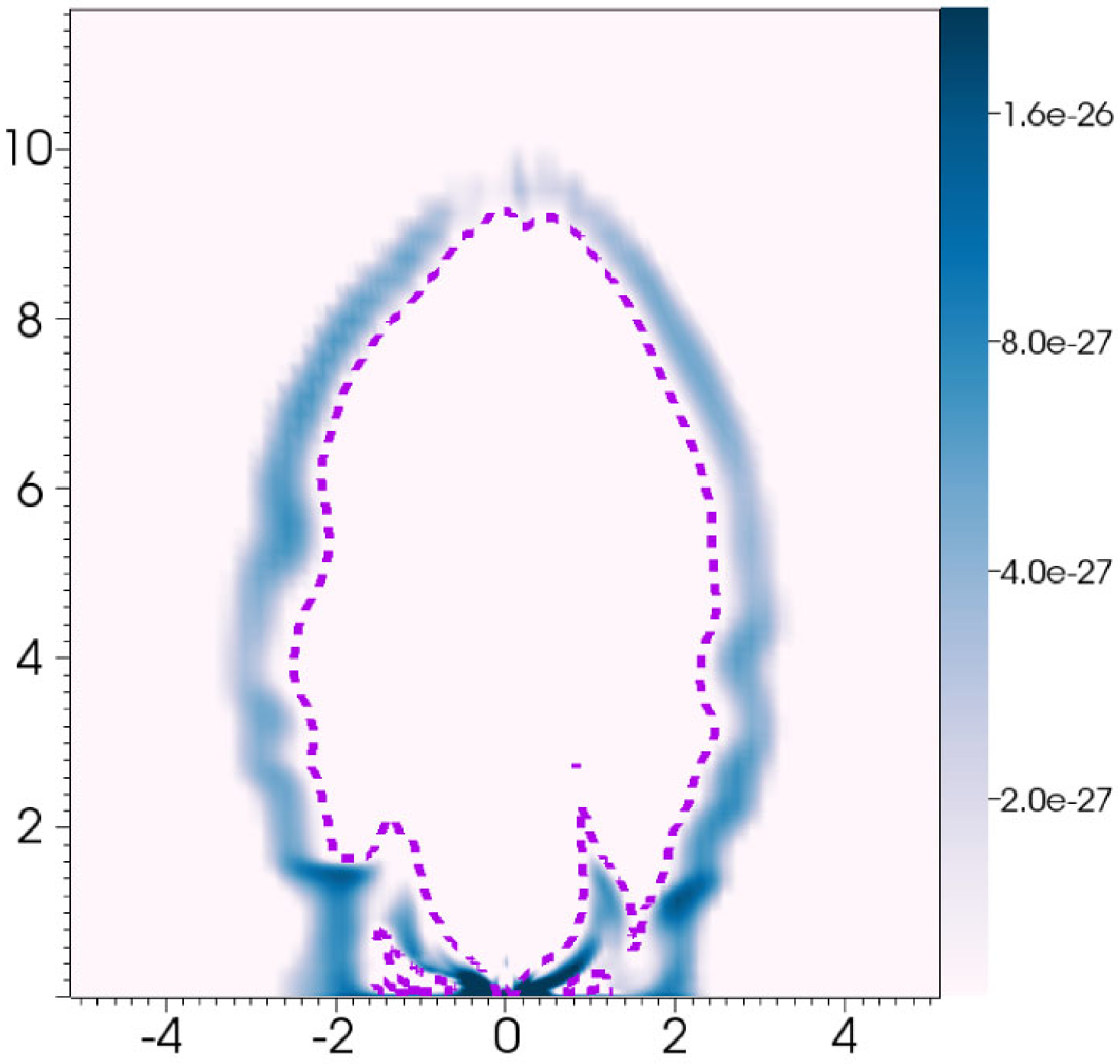}
 \includegraphics[width=0.3\textwidth]{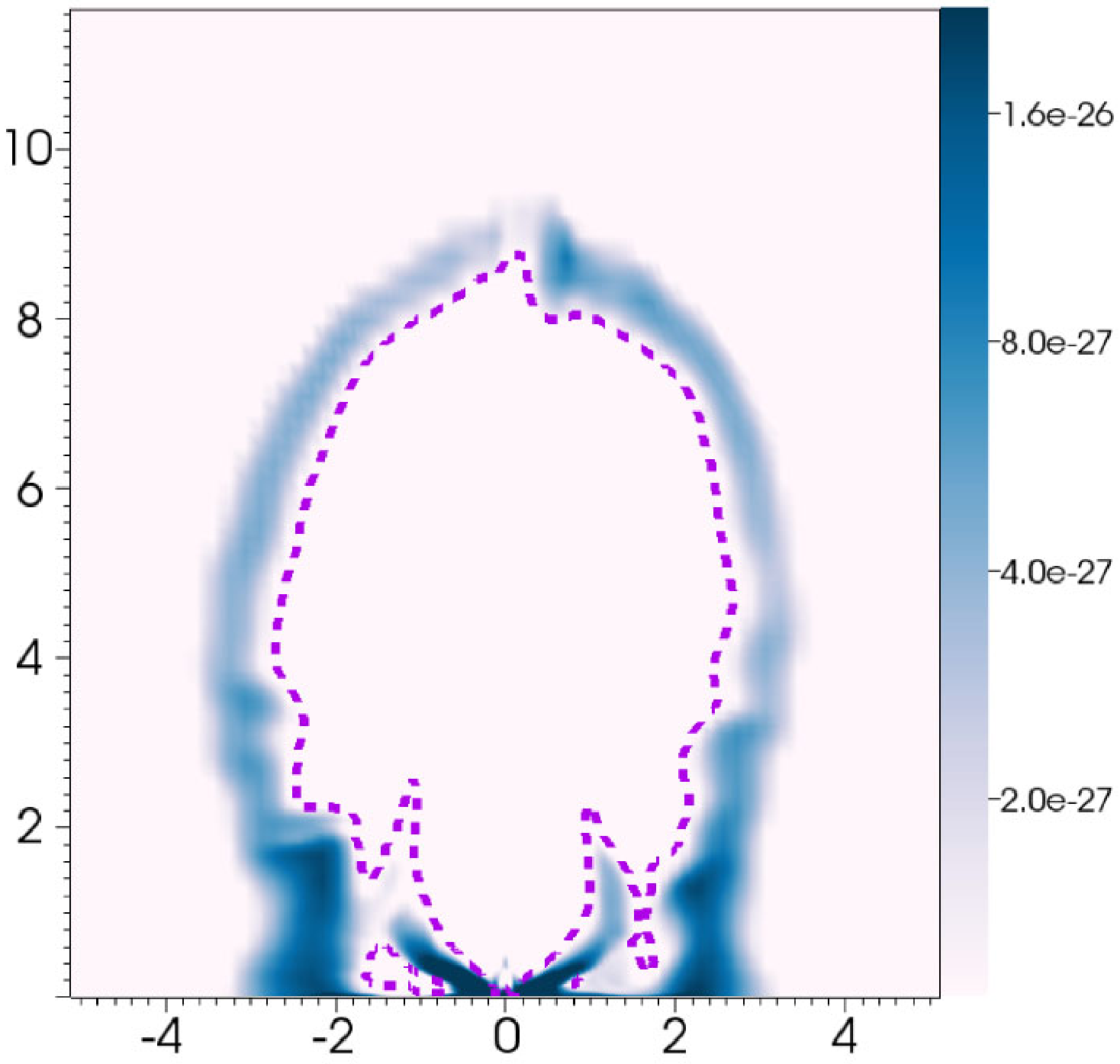}
 \includegraphics[width=0.3\textwidth]{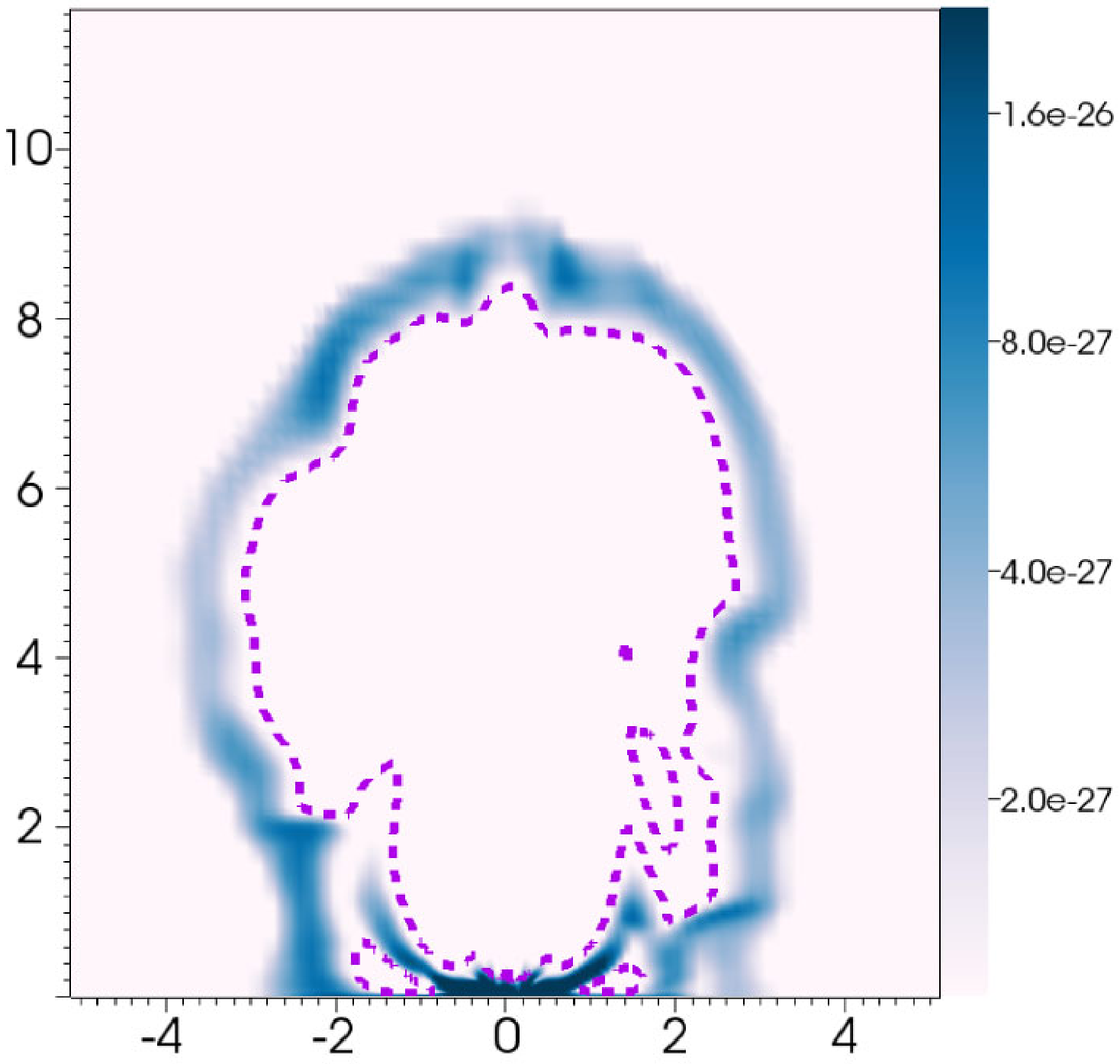}
 \vspace{-0.6cm}
 \end{center}
 \caption{Simulated slices of the gamma-ray volume emissivity in $E_{\gamma}\geq0.1$ GeV range. Here for simplicity, we only plot the contribution of $\pi^{0}$ decays. The three slices from left to right correspond to run A, B and E, respectively. The contour lines in each panels represent the density with the same meaning as in Figure \ref{plot1}. }
 \label{plot4}
\end{figure*}

 From Equation (\ref{Intensity}), our calculation result of run A is shown in Figure \ref{plot3}, in which the vertical axis shows $E^{2} I_{\gamma}$. The result is averaged within the scope of the \emph{Fermi} bubbles in the sky with latitudes larger than $10^{\circ}$.  Note that we have multiplied the intensity of $\pi^{0}$ decays (dashed line) by a factor 1.5 as a rough correction for heavier ions (\citealt{Mori1997}), i.e., the dashed line shows 1.5$E^{2} I_{\pi^{0}\rightarrow\gamma}$. Our model can  fit the latest observation  quite well. Specifically, the $\pi^{0}$ decays, the secondary leptons, and the primary CRe contribute 75\%, 7\% and 18\% of the total intensity, respectively. 
$\pi^{0}$ decays dominant the origin of gamma-rays at $E_{\gamma}\ga 0.3$ GeV; at $E_{\gamma} \la 1$ GeV, IC of  leptons is important. The leptons can significantly soften the spectrum at $E_{\gamma} \la 1$ GeV, but still the hardening of spectrum in this band is clear, which can be regarded as the characteristic signature of $\pi^{0}$-decay.

The volume emissivity $4\pi j_{\gamma}=E_{\gamma}dN_{\gamma}/dE_{\gamma}$ for run A, B and E is shown in Figure \ref{plot4}. For simplicity, we only plot the emissivity of $\pi^{0}$ decays in this figure. The dashed line shows the CD. It is obvious that the gamma-ray emissivity in the permeated zone outside the CD is much higher than that inside the CD. This is because the density of the shocked ISM is much higher outside the CD. The thickness of the permeated zone is typically only about 0.6--0.7 kpc. Almost all of gamma-ray photons that are produced in $\pi^{0}$ decays come from this thin shell.
 Therefore, strictly speaking, it is the edge of the permeated zone caused by the CR diffusion, rather than the CD, that is the edge of the \emph{Fermi} bubbles.

\subsection{The Gamma-Ray Surface Brightness}

\begin{figure}[!htb]
  \centering
  \plotone{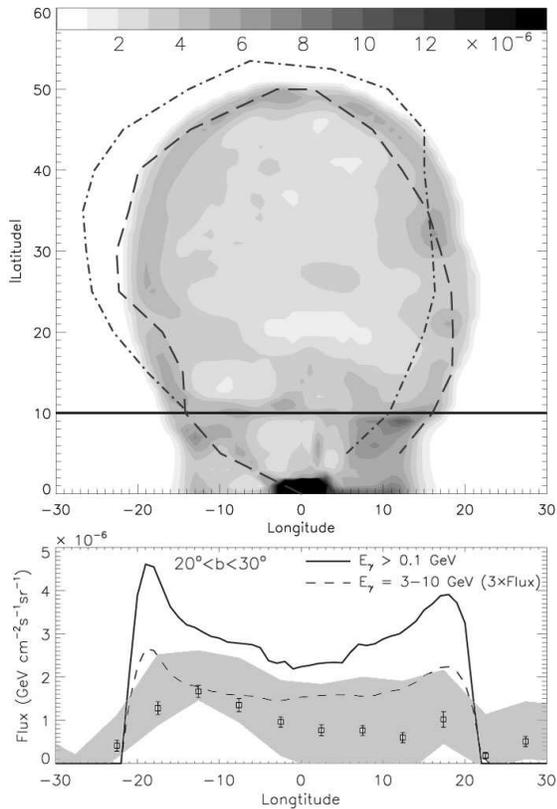}
   \vspace{-0.1cm}
   \caption{\emph{Top}: The gamma-ray intensity map for run A.
   \emph{Bottom}: Flux averaged between $b=20^{\circ}$ and $30^{\circ}$ along the longitude. The solid and dashed lines are our simulation results, while the rectangle with error bars and the shaded area are the observational results in 3--10 GeV band taken from  \citet{Ackermann2014}. The solid line is for $E_{\gamma}>0.1{\rm GeV}$ while the other lines and shaded area denote 3 times the flux. From the solid line, we can see the contrast between the bright edge and the dim center is about 1.6--2, and the boundary shows a width of about $3^{\circ}$.}
   \label{plot5}
\end{figure}

The gamma-ray surface brightness is shown  in Figure \ref{plot5}.
The contributions from both the $\pi^0$ decays and the IC (including primary and secondary leptons) are included, and the IC emissivity  is assumed to be constant inside the bubble.  The dashed and dot-dashed lines represent the edges of the observed north and south bubbles respectively (\citealt{Su10}). The intensity is integrated from 0.1 GeV to infinity, and is in units of ${\rm GeV~cm^{-2}~s^{-1}~sr^{-1}}$.  The observation by \citet{Ackermann2014} masks the Galactic plane at $|b|<10^{\circ}$, so we also focus on $|b|>10^{\circ}$.
The spatially averaged intensity for latitudes greater than $10^{\circ}$ is $3.3\times 10^{-6}~{\rm GeV~cm^{-2}~s^{-1}~sr^{-1}}$. From the figure we can find the following results.

\begin{itemize}
\item The simulated morphology is close to the observations, as in paper I.
\item The brightness does not vary with latitude, which is also consistent with observations.
\item The edge of the bubbles is not very sharp. Our simulation shows that the angular width of the edge of the bubble is $\sim 3^{\circ}$. This is  consistent with the observational result in \citet{Ackermann2014}. This angular width is caused by the projection effect of the permeated zone with a typical thickness of 0.6--0.7 kpc.
\item We find that the surface brightness of the bubbles is not very uniform, but shows a slight limb-brightened feature. The brightness near the edge is nearly twice the center (see the solid line in the bottom panel in Figure \ref{plot5}). This is because of the projection effect of the permeated zone. In the work of \citet{Ackermann2014}, the baseline model in 1--3 GeV and 3--10 GeV bands also shows a blurry limb-brightened feature around $b=30^{\circ}$ but with some uncertainties (refer to their Figures 22 and 23). To compare with the observational data, here we calculate our gamma-ray flux in 3--10 GeV. The result is shown in the bottom panel of Figure \ref{plot5} (dashed line). We can see that our result is roughly consistent with the data. Further observations are required to examine this issue.

\end{itemize}

 \citet{YangRZ2014} found that the morphology of the \emph{Fermi} bubbles is energy-dependent: they are more extended at high energies. From the figures in their paper, we estimate that the morphology at 10--30 GeV is about $3^{\circ}$, or equivalently 0.5 kpc, larger than at 1--2 GeV. Our model can explain this result. The physical reason is as follows. The above two gamma-ray bands roughly correspond to CRp with energy of $\sim$ 200 GeV and $\sim 20$ GeV, respectively.
In our model, the total diffusion duration of CRp is about 7 million yrs (i.e., the age of the {\it Fermi} bubbles). Combing this duration, the respective diffusion speed of CRp at the above two energies, and the inclination angle of the magnetic field, we can calculate the diffusion depth of the CRs with the above two energies. We find that their difference is $\sim 0.5$kpc.

\subsection{The Effects of Changing Parameters}


 In our fiducial model run A, we neglect the diffusion coefficient perpendicular  to the magnetic field, i.e., $\kappa_{\perp}=0$. In run B, we include this transverse diffusion (see Table 1). In this case, we find that the permeated zone becomes slightly thicker (see the middle panel of Figure \ref{plot4}), hence the width of the boundary of the bubble is slightly larger, and the contrast ratio of the bright edge and the dim center is smaller. The total luminosity is slightly enhanced (see Table \ref{table1}).


 In run C, we have tested a case with the ISM density reduced by a factor of 2. We find that  the gamma-ray intensity decreases by almost a factor of 3.


Run D tests the effect of changing the ``CR parameter''  $\eta_{CR}$. Comparing  runs A and D, we find that the final luminosity decreases by a factor of 2 when $\eta_{CR}$ decreases by one third.


Increasing the mass flux in the winds but keeping all other parameters unchanged will obviously increase the gamma-ray intensity produced. Now we discuss the effect of another parameter, the transition radius between the hot and cold accretion flows $R_{\rm tr}$. In our fiducial model, we set $R_{\rm tr}\sim 200 R_s$. If $R_{\rm tr}$ is larger, the mass flux in the winds will be larger according to Equation (1), but the total kinetic power of thermal winds does not change according to Equation (2). In run E we increase the mass flux of the winds by a factor of 4 compared to run A while keeping the kinetic power of thermal gas in the winds unchanged. We find that, the gamma-ray luminosity only slightly increases. The increase is because  more thermal nuclei are involved in the $pp$ collisions {\it inside} the bubble. The increase is small since the gamma-rays are mainly produced in the permeated zone, whose properties are not sensitive to the mass outflow rate but mainly to the power of winds.

\section{SUMMARY}

In a previous work (paper I) we have proposed an ``accretion wind'' model for the formation of the {\it Fermi} bubbles. Using hydrodynamic numerical simulations, we have shown in that work that the morphology of the bubbles can be successfully explained by the interaction between the winds launched from the hot accretion flow around the supermassive black hole Sgr A*. In the present work we continue this work by interpreting the gamma-ray emission of the bubbles. We invoke a hadronic model to explain the origin of gamma-ray radiation.  In this scenario, the gamma-ray photons are produced by the decay of neutral pions produced by the inelastic collisions between CRp and thermal nuclei in the ISM. The CRp are produced within and perhaps more importantly at the surface of the hot accretion flow by processes like magnetic reconnection. These CRp will be carried by winds launched from the accretion flow and interact with the thermal particles in the ISM in the interface between the winds and ISM.
The parameters of the winds are taken from the small scale MHD numerical simulations of hot accretion flow (\citealt{Yuan2015}). The distribution of these CRp is calculated in the present work by three dimensional MHD simulations, considering the advection and diffusion of CRp in the thermal winds when the magnetic field is present.  We then have calculated the emitted gamma-ray spectrum from the bubbles based on the simulation data. We have compared the results with observations and found that our model can not only explain the morphology of the bubbles but also explain the gamma-ray spectrum very well (Figure \ref{plot3}).

Some details of the results can be summarized below.

\begin{itemize}

\item Although the winds are almost isotropic, the CRp do not expand isotropically into the galactic halo. Rather, an ``umbrella-like'' structure is formed (Figure \ref{plot2}). The CRp first move along the ``handle'' of the ``umbrella'', after they reach the top of the bubbles, they diffuse downwards along the surface of the ``umbrella''. In this scenario, CRp occupy a relatively small volume, thus the total energy of CRp does not lose much. This kind of distribution is also conducive to form a constant surface brightness in the latitude direction.
\item We find that the magnetic field is slightly misaligned with the CD, which makes most of the CRp diffuse across the CD into some depth of the shocked ISM, forming a permeated zone (Figure \ref{plot1}). The thickness of the permeated zone is only about 0.6--0.7 kpc, but is the dominant region for generating the gamma-ray photons since thermal nuclei are much denser than inside the CD. Therefore it is the edge of the permeated zone, rather than the CD, that defines the edge of the \emph{Fermi} bubbles.

\item Our model can fit the observed gamma-ray spectrum quite well (Figure \ref{plot3}). The gamma-ray  mainly comes from the decay of neutral pions produced in the $pp$ collisions.

\item Our accretion wind model can explain the width of the boundary of the \emph{Fermi} bubbles, which is caused by the projection effect of the permeated zone (Figure \ref{plot5}).

\item Our result shows a somewhat limb-brightened surface brightness, with the contrast between the bright edge and the dim center of 1.6--2. This is roughly consistent with the observed blurry limb-brightened feature in 1--3 GeV and 3--10 GeV maps in \citet{Ackermann2014} (bottom panel of Figure \ref{plot5}).

\end{itemize}

\section{Discussion}

\subsection{The Giant Radio Loop Structure}

In paper I we have mentioned that our model can explain the {\it ROSAT} X-ray
features.
 Now we discuss the interpretation of another structure, the giant radio loop called Loop I. This structure was first observed in the radio band (\citealt{Large1962}), and most recently also detected in gamma-ray (\citealt{Ackermann2014}). Loop I radio image roughly surrounds the north \emph{Fermi} bubble, and the recent gamma-ray observation also shows that the gamma-ray image of Loop I happens to surround both the north \emph{Fermi} bubble and the lower part of the south \emph{Fermi} bubble (refer to Figure 13 in \citealt{Ackermann2014}). The gamma-ray spectral index is $\alpha \approx -2.4$, which is significantly softer than that of the \emph{Fermi} bubbles (\citealt{Su10}, \citealt{Ackermann2014}). The origin of Loop I is still not clear.

The North Polar Spur (NPS) which was observed in both the radio  and X-ray bands, is the brightest region in Loop I. \citet{Guo1} propose that this structure shares the same origin with the \emph{Fermi} bubbles and corresponds to the shocked ISM.
Here we focus on the gamma-ray emission of Loop I (or NPS).
The forward shock outside the \emph{Fermi} bubbles can accelerate both electrons and protons. The CRe could emit radio emission through synchrotron radiation. Gamma-ray emission could be produced through IC scattering of soft seed photons by CRe,  or pion decays through $pp$ collisions. Quantitatively, if the gamma ray emission is dominated by IC of CRe, the observed gamma-ray spectral index requires that the energy spectral index of CRe is roughly $p\approx -2.4$. If these electrons are mainly accelerated through \emph{Fermi} first-order acceleration in the forward shock, the required Mach number of the shock is $\sim 4$ (\citealt{Drury1983}). The Mach number in our simulation is $\sim 4-7$, close to the required value.

\subsection{Winds versus Jets}

Since the accretion flow around Sgr A* has been a hot accretion flow for millions of years, a jet must also exist together with the winds (Yuan \& Narayan 2014). In our ``accretion wind'' model of the {\it Fermi} bubbles, we do not include jet.  We assume that the role of jet is negligible to the formation of the bubbles. If we do not consider complexity such as the precession or turbulence of ISM, the jet will simply pierce through the ISM in a narrow low-density channel through which the jet can freely flow, without producing a bubble-like structure (\citealt{Vernaleo2006}).

 On the other hand, in the literature a type of jet model has been proposed to explain the formation of the {\it Fermi} bubbles (e.g., \citealt{Guo1, Guo2}; \citealt{Yang2012, Yang2013}). Several assumptions are adopted in this model. Firstly,
 to explain the bubbles which is perpendicular to the Galactic plane, the jet is required to be perpendicular to the Galactic plane as well. This assumption seems to be too strong since radio surveys show this is generally not the case (\citealt{Zubovas2011}). Secondly, the required mass lost rate in the jet is too large. It is $\sim 0.3 \dot{M}_{\rm Edd}$ in the hydrodynamical model (\citealt{Guo2}) and $\sim 300 \dot{M}_{\rm Edd}$ in the MHD model (\citealt{Yang2012, Yang2013}), respectively.  Theoretical models of jet formation predict that only a small fraction of the accretion matter can go into the jet, say $\sim 10\%$ for the ``disk jet'' or much smaller for the ``Blandford-Znajeck'' jet (see  \citealt{YuanNarayan2014} for the review of various jet models and Figure 5 in \citealt{Yuan2015} for the mass flux in the disk jet). As we state in \S2.1, the accretion rate in the underlying accretion flow in the past was likely only $\sim 10^{-2}\dot{M}_{\rm Edd}$. Thirdly,  the velocity of the jet required in the Yang et al. (2012, 2013) is very low, only $\sim 0.03c$. In \citet{Guo2} it is $\sim 0.1c$, which is still much smaller than the numerical simulation and some observational results, unless significant deceleration of jet occurs. For example, the speed of the jet in M87 is $\ga 0.986c$ (e.g., \citealt{Biretta1999}).

Our result has interesting implications to the formation of cavities and bubbles often observed in galaxy clusters. The general explanation is that they are formed because of  the interaction between jets and ISM. The usual argument is that we have observed a jet there. However, we now know that for a hot accretion flow, jets and winds are symbiotic (\citealt{YuanNarayan2014}). We don't see winds does not mean they do not exist, because winds are hard to be detected. Compared to jets, in the wind the magnetic field is much weaker thus synchrotron emission is weaker. Compared to the winds from a cold accretion flow, the temperature of winds from a hot accretion flow is much higher thus the gas is fully ionized. Therefore there will be almost no lines. By analogy with the formation of the {\it Fermi} bubbles, it will be interesting to investigate the possibility that cavities and bubbles are formed by winds rather than jets.

\acknowledgements
 We thank Roland M. Crocker, Yi-Lei Tang for very useful discussions. This project was supported in part by the National Basic Research Program of China (973 Program, grant 2014CB845800), the Strategic Priority Research Program ¡°The Emergence of Cosmological Structures¡± of CAS (grant XDB09000000), and NSF of China (grants 11133005, 11103061, and 11222328).  MYS acknowledges support from the China Scholarship Council (No.\ [2013]3009).
 Use of ZEUS-3D, developed by D. Clarke at the ICA ( http://www.ica.smu.ca) with support from NSERC, is acknowledged.
 This work made use of the High Performance Computing Resource in the Core Facility for Advanced Research Computing at Shanghai Astronomy Observatory.

\appendix

 In this part we introduce the three components of the GMF adopted in our simulation.

 {\em Toroidal halo component: $\vec B_{tor}$} (see \citealt{Jansson2012}). This component only has one direction: $\vec \phi$ in the north halo, or -$\vec \phi$ in the south halo. The strength of the field is in the form:
 \begin{gather}
  B_{tor}=B_{tor0}e^{-|z|/z_{0}}L(z,h_{disk},w_{disk})(1-L(r,r_{n},w_{h})), \label{Btor} \\
  L(z,h,w)=(1+e^{-2(|z|-h)/w})^{-1},
 \end{gather}
 in which $z_{0}=5.3$ kpc, $h_{disk}=0.4$ kpc, $w_{disk}=0.27$ kpc, $r_{n}=9.2$ kpc and $w_{h}=0.2$ kpc. We set the $B_{tor0}=1.4~\mu$G which is the value of northern halo. Considering that we have not included the disk magnetic field component, here we set $L(z,h_{disk},w_{disk})$ to be a constant of 1, which is a very good approximation for $z>0.7$ kpc, otherwise the toroidal magnetic field would be too weak near the galactic plane.

\begin{figure*}[!htb]
  \centering
  \begin{center}
   \includegraphics[width=0.30\textwidth]{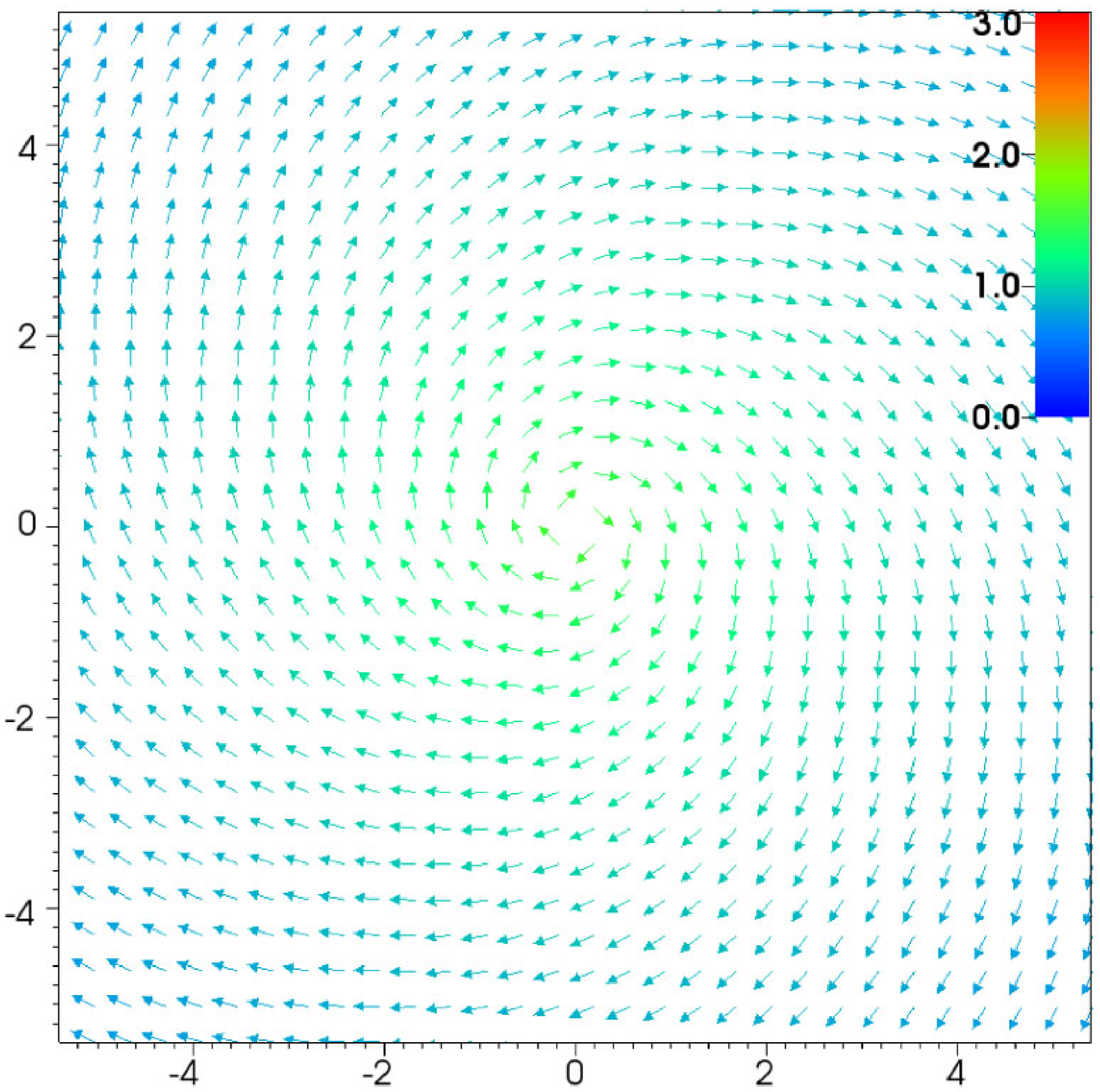}
   \includegraphics[width=0.293\textwidth]{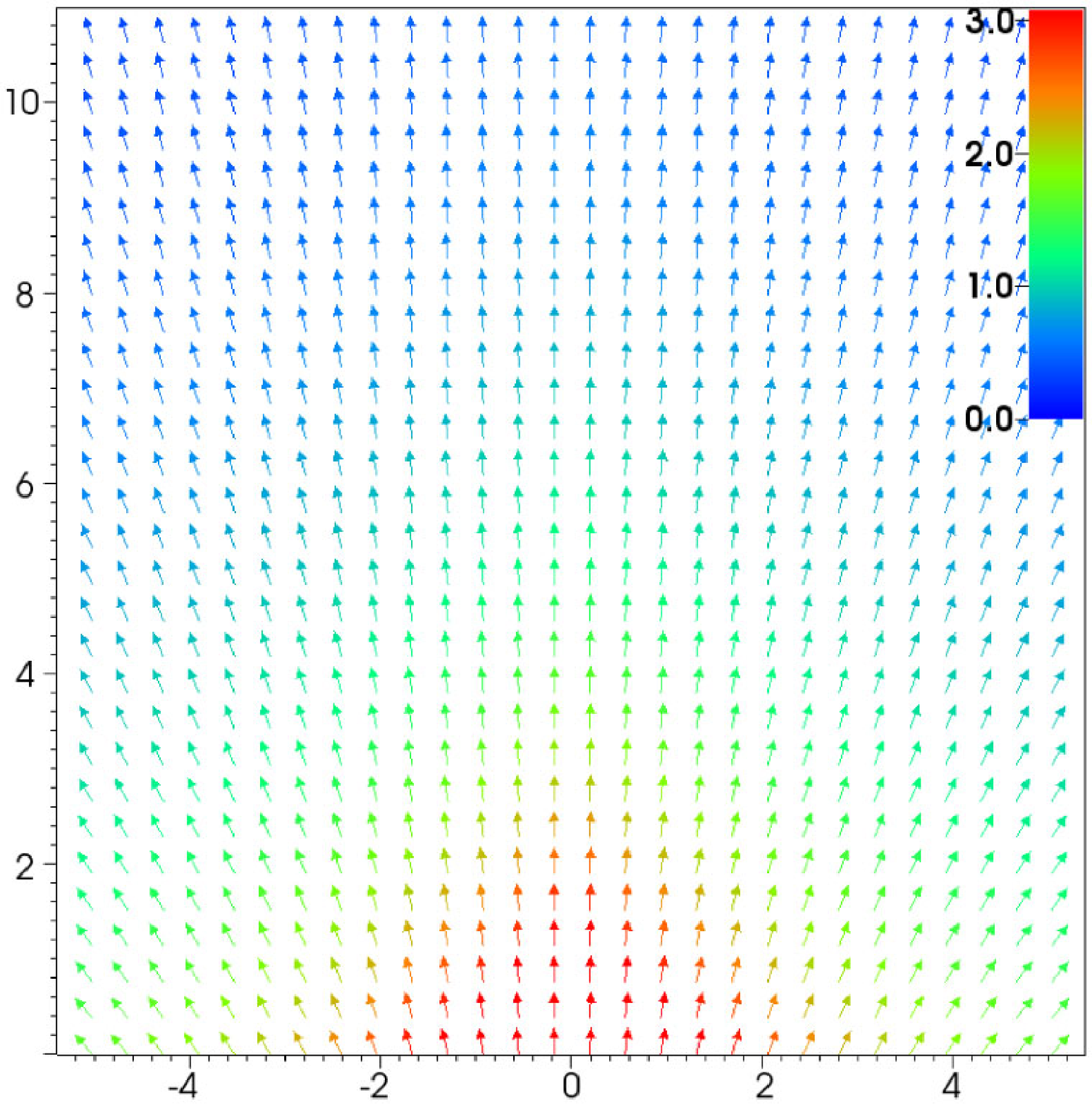}
   \includegraphics[width=0.30\textwidth]{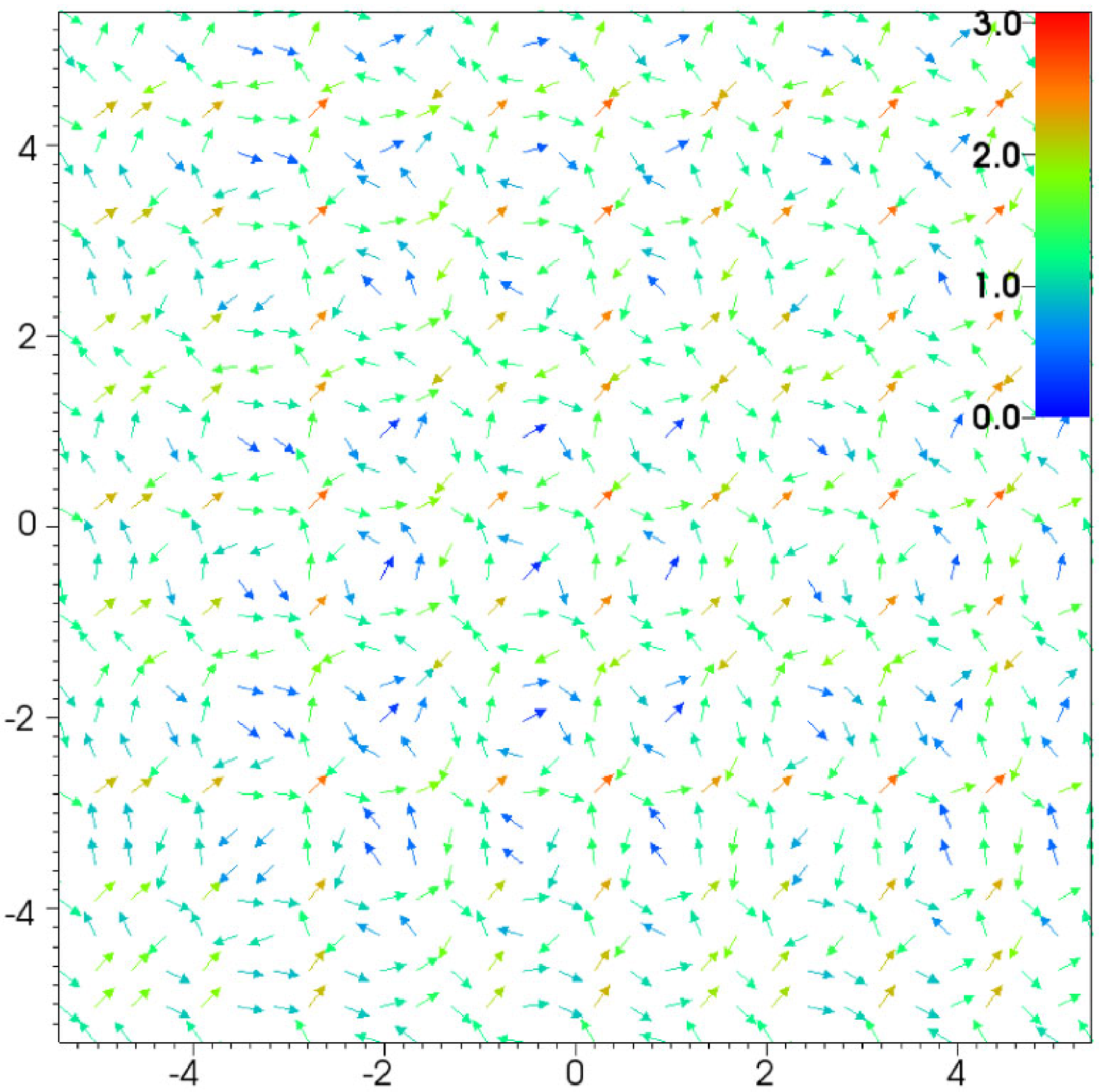} \\
  \end{center}
  \caption{Magnetic field configurations on different slices. \emph{Left}: purely large-scale regular magnetic field (without the ``tangled'' component) on the slice of $Z=4$ kpc. \emph{Middle}: purely large-scale regular magnetic field (without the ``tangled'' component) on the slice of $Y=0$ kpc. \emph{Right}: purely tangled magnetic field on the slice of $Z=4$ kpc. Color bars show the strength of magnetic field in units of $\mu$G.}
 \label{plot0}
\end{figure*}

 {\em Out-of-plane component: $\vec B_{X}$} (see \citealt{Jansson2012}).
 The field lines of the out-of-plane component are straight lines in the space, and show a mirror symmetry about the $Z=0$ plane (see the middle panel in Figure \ref{plot0} for this magnetic field). This magnetic field lines look like ``X-shaped'', and is similar to the ``X-shaped'' field lines in some edge-on galaxies, such as NGC 891, NGC 4666, NGC 5775 and so on (\citealt{Haverkorn2012}). If we set the cylindrical radius where the out-of-plane field line intersecting the plane of $Z=0$ to be $R=r_{p}$, the magnetic field strength on the $Z=0$ plane is \be b_{X}(r_{p})=B_{X0}e^{-r_{p}/r_{X}},  \ee where $B_{X0}$ is $4.4~\mu$G, and $r_{X}$ is 2.9 kpc.
 The space can be divided into two parts, divided by a critical cylindrical radius $R=r^{c}_{X}$ on $Z=0$ plane and a constant elevation angle of $49^{\circ}$ starting from this critical circle. The elevation angles are different in the two parts. Specially, for $r_{p}\geq r^{c}_{X}$, the elevation angle is a constant of $49^{\circ}$ , which is represented by $\theta^{0}_{X}$. The magnitude of out-of-plane magnetic field in $r_{p}\geq r^{c}_{X}$ region is described by:
 \begin{gather}
 |\vec B_{X}|=b_{X}(r_{p}) \frac{r_{p}}{r}, \label{BX1} \\
 r_{p}=r-\frac{|z|}{\tan \theta^{0}_{X}},
 \end{gather}
 where $r^{c}_{X}$ is a constant of 4.8 kpc.

 For $r_{p} < r^{c}_{X}$, the elevation angle linearly changes from $\theta^{0}_{X}$ at $R=r_{p}$ to 0 at $R=0$. The magnetic field is described in the form:
 \begin{gather}
 |\vec B_{X}|=b_{X}(r_{p})(\frac{r_{p}}{r})^2, \label{BX2} \\
 r_{p}=\frac{r\cdot r^{c}_{X}}{r^{c}_{X}+|z|/\tan \theta^{0}_{X}}, \\
 \theta_{X}(r,z)=\arctan(\frac{|z|}{r-r_{p}}).
 \end{gather}

 {\em Tangled field: $\vec B_{tb}$}. The direction of the magnetic field $\vec B_{tb}$ is set to be a spatially periodic form to mimic the ``tangled'' status, in the form of:
\begin{gather}
 B_{1}=B_{0}[\sin(K_{2}y+\Delta_{2})+\sin(K_{3}z+\Delta_{3})], \label{B1} \\
 B_{2}=B_{0}[\sin(K_{1}x+\Delta_{1})+\sin(K_{3}z+\Delta_{3})], \label{B2}\\
 B_{3}=B_{0}[\sin(K_{1}x+\Delta_{1})+\sin(K_{2}y+\Delta_{2})], \label{B3}
\end{gather}
 where $B_{0}$ is a constant of 0.5 $\mu$G, $K_{1,2,3}$ are set to be $6.28~\rm kpc^{-1}$ which means that the initial GMF is ``tangled'' on length scale of 1.0 kpc in each direction, and the phases $\Delta_{1,2,3}$ are set to be 0.5, 0.5 and 1.0, respectively. To explore the influence of the phases, we also try another set of data: $\Delta_{1,2,3}=$ 1.5, 1.5, and 1.0, respectively. In this case, the magnetic field on the galactic polar axis is relatively stronger, but we find that the final result is almost the same.
 It's easy to prove that magnetic field with this form satisfies the zero divergence.

 The three components are shown in Figure \ref{plot0}. The large-scale regular magnetic field is the sum of $\vec B_{tor}$ and $\vec B_{X}$.

\end{document}